\documentclass[aps,pre,reprint,superscriptaddress,nofootinbib]{revtex4-1}
\usepackage{graphicx}
\usepackage{amssymb}
\usepackage {amsmath} 

\begin{document}


\title{Constructing a class of topological solitons in magnetohydrodynamics}

\author{Amy Thompson}
\affiliation{\small Dept. of Physics, University of California, Santa Barbara, CA 93106}
\affiliation{\small Huygens Laboratory, Leiden University, PO Box 9504, 2300 RA Leiden, The Netherlands}
\author{Joe Swearngin}
\affiliation{\small Huygens Laboratory, Leiden University, PO Box 9504, 2300 RA Leiden, The Netherlands}
\author{Alexander Wickes}
\affiliation{\small Huygens Laboratory, Leiden University, PO Box 9504, 2300 RA Leiden, The Netherlands}
\author{Dirk Bouwmeester}
\affiliation{\small Dept. of Physics, University of California, Santa Barbara, CA 93106}
\affiliation{\small Huygens Laboratory, Leiden University, PO Box 9504, 2300 RA Leiden, The Netherlands}


\date{\today}

\begin{abstract}
We present a class of topological plasma configurations characterized by their toroidal and poloidal winding numbers, $n_t$ and $n_p$ respectively. The special case of $n_t=1$ and $n_p=1$ corresponds to the Kamchatnov-Hopf soliton, a magnetic field configuration everywhere tangent to the fibers of a Hopf fibration so that the field lines are circular, linked exactly once, and form the surfaces of nested tori. We show that for $n_t \in \mathbb{Z}^+$ and $n_p=1$ these configurations represent stable, localized solutions to the magnetohydrodynamic equations for an ideal incompressible fluid with infinite conductivity. Furthermore, we extend our stability analysis by considering a plasma with finite conductivity and estimate the soliton lifetime in such a medium as a function of the toroidal winding number. 
\end{abstract}

\pacs{}

\maketitle



%




\section{Introduction}

A \emph{hopfion} is a field configuration whose topology is derived from the Hopf fibration. The Hopf fibration is a map from the 3-sphere ($\mathbb{S}^3$) to the 2-sphere ($\mathbb{S}^2$) such that great circles on $\mathbb{S}^3$ map to single points on $\mathbb{S}^2$. The circles on $\mathbb{S}^3$ are called the \emph{fibers} of the map, and when projected stereographically onto $\mathbb{R}^3$ the fibers correspond to linked circles that lie on nested, toroidal surfaces and fill all of space. The fibers can be physically interpreted as the field lines of the configuration, giving the hopfion fields their distinctive toroidal structure \cite{Irvine2008}. 

Hopfions have been shown to represent localized topological solitons in many areas of physics -  as a model for particles in classical field theory \cite{Skyrme1962}, fermionic solitons in superconductors \cite{Ran}, particle-like solitons in superfluid-He \cite{Volovik1977}, knot-like solitons in spinor Bose-Einstein (BE) condensates \cite{Kawaguchi2008} and ferromagnetic materials \cite{Dzyloshinskii1979}, and topological solitons in magnetohydrodynamics (MHD) \cite{Kamchatnov1982}. The Hopf fibration can also be used in the construction of finite-energy radiative solutions to Maxwell's equations and linearized Einstein's equations \cite{Swearngin2013}. Some examples are Ra{\~n}ada's null EM hopfion \cite{Ranada1989,Ranada2002} and its generalization to torus knots \cite{Irvine2008, Irvine2013, Kobayashi2013}.

Topological solitons are metastable states. They are not in an equilibrium, or lowest energy, state, but are shielded from decay by a conserved topological quantity. The energy $E$ is a function of a scale factor, typically the size $R$ of the soliton, so that the field could decrease its energy by changing this parameter. However, the topological invariant fixes the length scale and thus the energy. In condensed states (superconductors, superfluids, BE condensates, and ferromagnets) the topological structure is physically manifested in the order parameter, which is associated to a topological invariant. For example, the hopfion solutions in ferromagnets are such that the Hopf fibers correspond to the integral curves of the magnetization vector $\vec m$. The associated Hopf invariant is equal to the linking number of the integral curves of $\vec m$. 

For many systems the solution can still decay by a continuous deformation while conserving the topological invariant. Another physical stabilization mechanism is needed to inhibit collapse \cite{Kalinkin2007}. For example, this can be achieved for superconductors with localized modes of a fermionic field \cite{Pismen2002}, for superfluids by linear momentum conservation \cite{Volovik1977}, for BE condensates with a phase separation from a second condensate \cite{Battye2002}, and for ferromagnets with conservation of the spin projection $S_z$ \cite{Zhmudskii1999}. 

In MHD, the topological structure is present in the magnetic field. The topological soliton of Kamchatnov has a magnetic field everywhere tangent to a Hopf fibration, so that the integral curves of the magnetic field lie on nested tori and form closed circles that are linked exactly once.  The Hopf invariant is equal to the linking number of the integral curves of the magnetic field, which is proportional to the magnetic helicity. In addition to the topological invariant, another conserved quantity is required. MHD solitons can be stabilized if the magnetic field has a specific angular momentum configuration which will be discussed below.

Because of the importance of topology in plasma dynamics, there has previously been interest in generalizing the Kamchatnov-Hopf soliton \cite{Semenov2002}. The topology of field lines has been shown to be related to stability of flux tube configurations, with the helicity placing constraints on the relaxation of magnetic fields in plasma \cite{Candelaresi2012, Candelaresi2011influence}. Magnetic helicity gives a measure of the structure of a magnetic field, including properties such as twisting, kinking, knotting, and linking \cite{Berger1984, Berger1999}. Simulations have shown that magnetic flux tubes with linking possess a longer decay time than similar configurations with zero linking number \cite{DelSordo2010, Candelaresi2011helical, Candelaresi2011trefoil}.  Recently, higher order topological invariants have been shown to place additional constraints on the evolution of the system \cite{Candelaresi2012, Yeates2010, Yeates2011}. The work presented in this paper distinguishes itself from these topological studies of discrete flux tubes in the sense that we are considering the topology and stability of continuous, space-filling magnetic field distributions. Furthermore, our results are analytic, rather than based on numerical simulations.

\begin{figure*}[t]
\begin{center}
\includegraphics{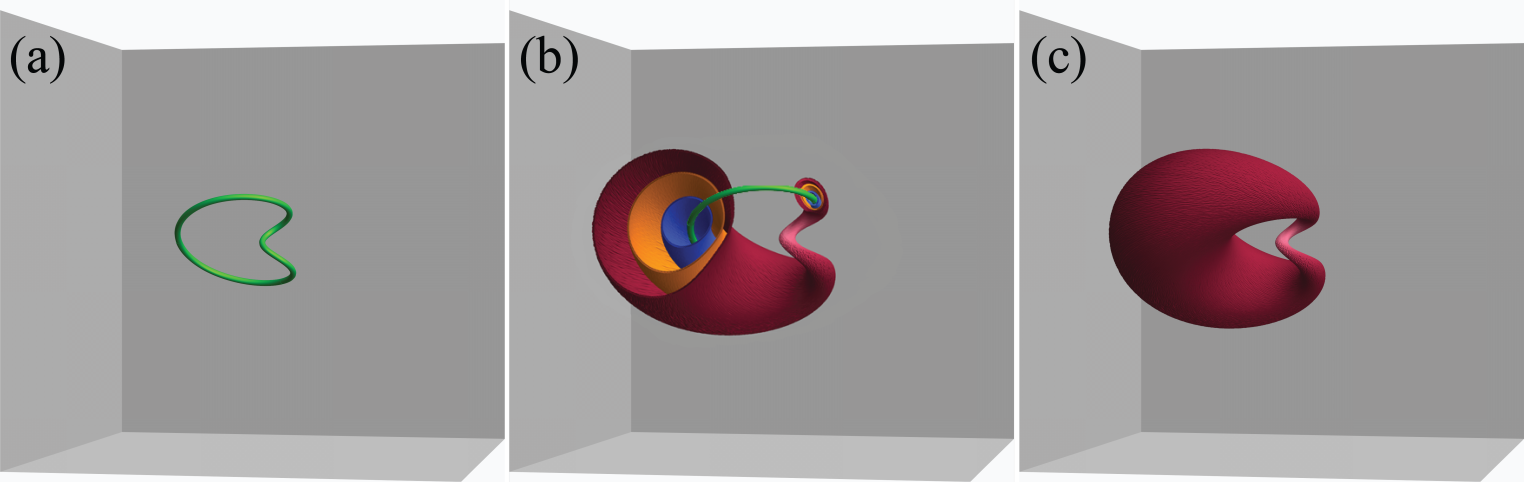}
\caption{(Color) One lobe of the field configuration for $n_{p}=1$ and $n_{t}=2$. (a) A single, closed core magnetic field line. (b) The core field line is surrounded by nested toroidal surfaces, shown in cross section. (c) A complete magnetic surface filled entirely by one field line. \label{fig:surfaces}}
\end{center}
\end{figure*}

There are many applications where magnetic field topology has a significant effect on the stability and dynamics of plasma systems. For example, toroidal magnetic fields increase confinement in fusion reactors \cite{White2001book,Haeseleer1991book}, and solving for the behavior of some magnetic confinement systems is only tractable in a coordinate system based on a known parameterization of the nested magnetic surface topology \cite{Boozer1982,Hamada1962,Haeseleer1991book}. In astrophysics, the ratio of the toroidal and poloidal winding of the internal magnetic fields impacts many properties of stars, including the shape \cite{Chandrasekhar2013, Wentzel1961} and momentum of inertia \cite{Mestel1981},
as well as the gravity wave signatures \cite{Lasky2013} and disk accretion \cite{Ghosh1978} of neutron stars. The new class of stable, analytic MHD solutions presented in this paper may be of use in the study of fusion reactions, stellar magnetic fields, and plasma dynamics in general.

The MHD topological soliton is intimately related to the radiative EM hopfion solution. The EM hopfion constructed by Ra{\~n}ada is a null EM solution with the property that the electric, magnetic, and Poynting vector fields are tangent to three orthogonal Hopf fibrations at $t=0$. The electric and magnetic fields deform under time evolution, but their field lines remain closed and linked with linking number one. The Hopf structure of the Poynting vector propagates at the speed of light \emph{without deformation}. The EM hopfion has been generalized to a set of null radiative fields based on torus knots with an identical Poynting vector structure \cite{Irvine2013}. The electric and magnetic fields of these toroidal solutions have integral curves that are not single rings, but rather each field line fills out the surface of a torus.

The time-independent magnetic field of the topological soliton is the magnetic field of the radiative EM hopfion at $t=0$
\begin{equation}
\mathbf{B}_{\rm soliton}(\mathbf{{x}}) = \mathbf{B}_{\rm hopfion}(t=0,\mathbf{{x}}).
\label{Bsoliton}
\end{equation}
The soliton field is then sourced by a stationary current
\begin{equation}
\mathbf{j}(\mathbf{{x}}) = \frac{1}{\mu_0} \mathbf{\nabla}  \times \mathbf{B}_{ \rm soliton}(\mathbf{{x}}).
\label{jsoliton}
\end{equation}
We will use this relationship, along with the generalization of the EM hopfion to toroidal fields of higher linking number, in order to generalize the Kamchatnov-Hopf topological soliton to a class of stable topological solitons in MHD. We will also discuss how the helicity and angular momentum relate to the stability of these topological solitons.
  
\section{Generalization of the Kamchatnov-Hopf Soliton}

\begin{figure*}
\begin{center}
\includegraphics{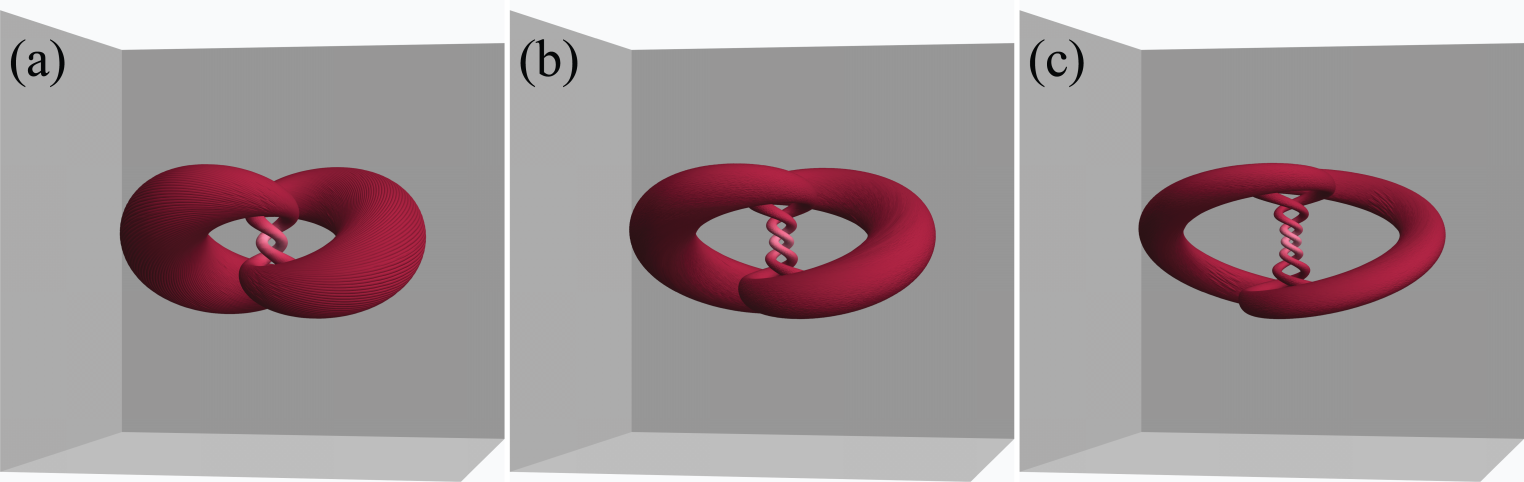}
\caption{(Color) Topological solitons in MHD with $n_{p}=1$ and (a)$n_{t}=2$, (b)$n_{t}=3$, and (c)$n_{t}=4$. A single magnetic field line fills out each of the linked, toroidal surfaces. \label{fig:toroidal}}
\end{center}
\end{figure*}

We construct the generalized topological soliton fields using Eqns. (\ref{Bsoliton}) and (\ref{jsoliton}) applied to the null radiative torus knots. The time-independent magnetic field of the soliton is identical to the magnetic field of the radiative torus knots at $t=0$. The magnetic field is sourced by a current, resulting in a stationary solution. 

The torus knots are constructed from the Euler potentials:
\begin{align}
\label{alpha}
\alpha =& \frac{(r^2 -t^2- 1) + 2 \imath z}{r^2-(t-\imath)^2}\\
\beta =& \frac{2 (x - \imath y)}{r^2-(t-\imath)^2}.
\label{beta}
\end{align}
where $r^2=x^2+y^2+z^2$. As Ref. \cite{Irvine2013} points out, at $t=0$ these are the stereographic projection coordinates on $\mathbb{S}^3$. The magnetic field of the torus knots is obtained from the Euler potentials for the Riemann-Silberstein vector ${\mathbf{F} = \mathbf{E} + \imath \mathbf{B}}$.\footnote{Note that the Riemann-Silberstein construction is a non-standard use of Euler potentials. We are following the method in Ref. \cite{Irvine2013}.}

 The solitons are found by taking the magnetic field of the torus knots at $t=0$
\begin{align}
\label{Bfield}
\mathbf{B} = Im[\nabla  \alpha^{n_{t}} \times \nabla \beta^{n_{p}}]\mid_{t=0}.
\end{align}

Each $(n_{t},n_{p})$ with $n_t,n_p=1,2,3...$ represents a solution to Maxwell's equations. A single magnetic field line fills the entire surface of a torus. These tori are nested and each degenerates down to a closed core field line that winds $n_t$ times around the toroidal direction and $n_p$ times around the poloidal direction, as illustrated in Fig. \ref{fig:surfaces}. A complete solution for a given $(n_t,n_p)$  is composed of pairs of these nested surfaces that are linked and fill all of space as shown in Fig. \ref{fig:toroidal}. For $n_g=gcd(n_{t}, n_{p})$, the solution is a magnetic field with $2n_g$ linked core field lines (knotted if $n_t>1$ and $n_p>1$). If $n_{t}=1$ and $n_{p}=1$, the solution is the Kamchatnov-Hopf soliton. We will analyze these fields and how the linking of field lines affects the stability of magnetic fields in plasma. In particular, for $n_{p}=1$ and $n_t \in \mathbb{Z}^+$, we will show that these fields can be used to construct a new class of stable topological solitons in ideal MHD. The solutions with $n_p \neq 1$ are not solitons in plasma, and their instability will be discussed in section \ref{sec:instability}.

\section{Stability Analysis}

In this section we assume the plasma is an ideal, perfectly conducting, incompressible fluid. In a fluid with finite conductivity, the magnetic field energy diffuses. Under this condition, one can estimate the lifetime of the soliton as will be shown in section \ref{sec:lifetime}.
 
First we consider the case where the poloidal winding number $n_{p}=1$ and the toroidal winding number $n_{t}$ is any positive integer. These will be shown to represent stable topological solitons in ideal MHD. In the next section, we will consider the solutions with $n_{p} \neq 1$. Using the method in this paper, these do not represent stable solitons, and we will discuss how this instability relates to the angular momentum.
 
To analyze the stability of these solutions, following the stability analysis in Ref.\cite{Kamchatnov1982}\footnote{Note that Ref. \protect{\cite{Kamchatnov1982}} uses CGS units and we use SI units in our analysis. The reference also has a typo - Eqn. (45) should have a factor of $R^2$ instead of $R$.}, we study the two scaled quantities of the system - the length scale $R$ which corresponds to the size of the soliton and $B_0$ which is the magnetic field strength at the origin. (The length scale R is also the radius of the sphere $\mathbb{S}^3$ before stereographic projection.) First we change to dimensionful coordinates by taking 
\begin{align}
\{ x,y,z \} \rightarrow & \{ \tfrac{x}{R},\tfrac{y}{R},\tfrac{z}{R} \} \\
| \mathbf{B}\left( 0,0,0 \right)| = & B_0.
\end{align}
The stability depends on three quantities - energy, magnetic helicity, and angular momentum - which are functions of $R$ and $B_0$. For a perfectly conducting plasma, the magnetic helicity $h_m$ is an integral of motion and is thus conserved. The magnetic helicity is also a topological invariant proportional to the linking number of the magnetic field lines. If the field can evolve into a lower energy state by a continuous deformation (therefore preserving the topological invariant) then it will be unstable. However, we will show that such a deformation does not exist because the angular momentum $M$ is also conserved and serves to inhibit the spreading of the soliton. 

\begin{figure*}
\begin{center}
\includegraphics{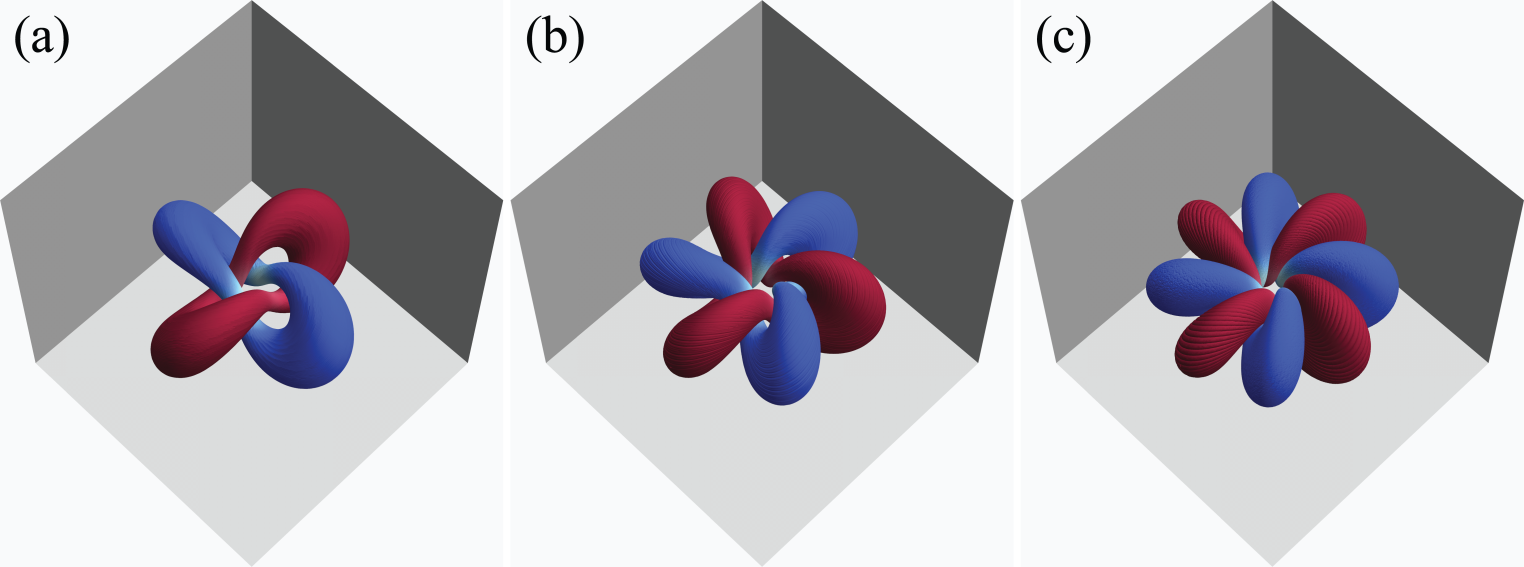}
\caption{(Color) The magnetic surfaces for $n_{t}=1$ and (a)$n_{p}=2$, (b)$n_{p}=3$, and (c)$n_{p}=4$. Solutions with $n_{p} \neq 1$ have zero angular momentum and are therefore not stable solitons. The magnetic field lines in each lobe wind in opposite directions, represented by the red and blue surfaces. \label{fig:poloidal}}
\end{center}
\end{figure*}

The magnetic helicity is defined as
\begin{align}
h_m =& \int{\mathbf{A} \cdot \mathbf{B} d^3x} 
\end{align}
where $\mathbf{A} =Im[\alpha^{n_t} \nabla \beta^{n_p}]$ is the vector potential. From Eqns. \eqref{alpha}-\eqref{Bfield}, it follows that
\begin{align}
h_m =& \frac{2n_{t}}{(n_{t}+1)} \pi^2 B_0^2 R^4.
\label{soliton_helicity}
\end{align}
The MHD equations for stationary flow are satisfied for a fluid with velocity 
\begin{align}
\mathbf{v} = \pm \frac{\mathbf{B}}{(\mu_0 \rho)^{\frac{1}{2}}}.
\end{align}
The energy of the soliton is given by
\begin{align}
E =& \int{\left( \frac{\rho v^2}{2}+ \frac{B^2 }{2 \mu_0} \right) d^3x} \\
 =& \int{ \frac{B^2}{\mu_0} d^3x}
\notag
\end{align}
so that
\begin{align}
E =& \frac{2 n_{t} \pi^2}{\mu_0}  B_0^2 R^3 \\
\propto & \frac{h_m}{R}.
\notag
\end{align}
The angular momentum is
\begin{align}
\mathbf{M} =& \rho \int{[\mathbf{x} \times \mathbf{v}] d^3x} \\
 =& \left(\frac{\rho}{\mu_0}\right)^{1/2} 4 n_{t} \pi^2 B_0 R^4 \hat{y}
\notag
\end{align}
where we took the positive velocity solution. We find that the conserved quantities $h_m$ and $\mathbf{M}$ fix the values of $R$ and $B_0$, 
\begin{align}
R =& \left(\frac{1}{8\pi^2 n_{t}(n_{t}+1)} \left(\frac{\mu_0}{\rho}\right)\frac{|M|^2}{h_m}\right)^{\frac{1}{4}},\\
 B_0 =& 2 n_{t}(n_{t}+1)\left(\frac{\rho}{\mu_0}\right)^{1/2} \frac{h_m}{|M|},
\notag
\end{align}
thus inhibiting energy dissipation. This shows that the solution given in Eqns. \eqref{alpha}-\eqref{Bfield} (and shown in Fig. \ref{fig:toroidal}) represents a class of topological solitons characterized by the parameter $n_{t} \in \mathbb{Z}^+$ for $n_{p}=1$. 

\subsection{Angular Momentum and Instability for $n_p \neq 1$ \label{sec:instability}}

For $n_{p} \neq 1$, the angular momentum for all $n_{t}$ is zero. Some examples of fields with $n_{t}=1$ and different $n_{p}$ values are shown in Fig. \ref{fig:poloidal}. The field lines fill two sets of linked surfaces. For a given pair of linked surfaces, the field in each lobe wraps around the surface in opposite directions. In Fig. \ref{fig:poloidal} the red and blue surfaces wind in opposite directions. This means that the contribution to the angular momentum of the two field lines cancels. In this case the length scale is not fixed by the conserved quantities. The energy can therefore decrease by increasing the radius and the fields are not solitons. 

\section{Finite Conductivity and Soliton Lifetime \label{sec:lifetime}}

To include losses due to diffusion, we need to consider a plasma with finite conductivity. We can estimate the soliton lifetime by dividing the energy by $dE/dt$, calculated before any energy dissipation \cite{Kamchatnov1982}. Since this is the maximum rate of energy dissipation, we can obtain a lower bound on the time it takes for the total energy to dissipate. Thus,
\begin{align}
\frac{dE}{dt} =& \frac{1}{\sigma}  \int j^2 d^3x \\
=& (3 n_{t} + 7 n_{t}^2 + 5 n_{t}^3) \frac{\pi^2 B_0^2 R}{\mu_0^2 \sigma}.
\end{align}
The resulting lifetime is
\begin{equation}
t_{n_{t}} \geq \frac{3 n_{t}}{3 n_{t} + 7 n_{t}^2 + 5 n_{t}^3} \mu_0 \sigma R^2.
\end{equation}
For higher $n_t$, the lifetime decreases although the helicity in Eqn. \eqref{soliton_helicity} increases. This result is interesting as we would have expected from the results regarding flux tubes mentioned previously that the lifetime would increase with increasing helicity.

\section{Conclusion} 

We have shown how to construct a new class of topological solitons in plasma. The solitons consist of two linked core field lines surrounded by nested tori that fill all of space. The solutions are characterized by the toroidal winding number of the core field lines and have poloidal winding number one in order to have non-zero angular momentum. We have shown that the conservation of linking number and angular momentum give stability to the solitons in the ideal case. For a plasma with finite conductivity, we have estimated the lifetime of the solitons and found that the lifetime decreases with increasing helicity. Finally, we note that there may be related generalizations of the hopfion fields in other physical systems, such as superfluids, Bose-Einstein condensates, and ferromagnetic materials.

\begin{acknowledgments}
The authors would like to acknowledge discussions with J.W. Dalhuisen and C.B. Smiet. This work is supported by NWO VICI 680-47-604 and NSF Award PHY-1206118.
\end{acknowledgments}


%

\end{document}